\documentclass[english]{article}
\usepackage[T1]{fontenc}
\usepackage[latin1]{inputenc}
\usepackage{amssymb}

\makeatletter

\newcommand{\noun}[1]{\textsc{#1}}

\newcommand{\lyxaddress}[1]{
\par {\raggedright #1
\vspace{1.4em}
\noindent\par}
}

\makeatletter

\date{}
\newcommand{\esp}{\vspace{.1cm}}
\newcommand{\hsp}{\hspace{1cm}}

\newcommand{\be}{\begin{equation}}
\newcommand{\ee}{\end{equation}}

\makeatother

\usepackage{babel}
\makeatother

\begin{document}

\title{Third Bose Fugacity Coefficient in One Dimension, \\
 As a Function of Asymptotic Quantities}

\author{A. Amaya-Tapia$^{1}$, S. Y. Larsen$^{2}$and M. Lassaut$^{3}$}

\maketitle

\lyxaddress{$^{1}$Instituto de Ciencias F\'{i}sicas, Universidad Nacional Aut\'onoma
de M\'exico AP 48-3, Cuernavaca, Mor. 62251, M\'exico. $^{2}$Department
of Physics, Temple University, Philadelphia, PA 19122, USA. $^{3}$Institut 
de Physique Nucl\'{e}aire, IN2P3-CNRS, Universit\'e Paris-Sud 11, F-91406 
Orsay Cedex, France.}

\begin{abstract}
In one of the very few exact quantum mechanical calculations of fugacity
coefficients, Dodd and Gibbs (\textit{J. Math.Phys}.,\textbf{15},
41 (1974)) obtained $b_{2}$ and $b_{3}$ for a one dimensional Bose
gas, subject to repulsive delta-function interactions, by direct integration
of the wave functions. For $b_{2}$, we have shown (\textit{Mol. Phys}.,\textbf{103},
1301 (2005)) that Dodd and Gibbs' result can be obtained from a phase
shift formalism, if one also includes the contribution of oscillating
terms, usually contributing only in 1 dimension. Now, we develop an
exact expression for $b_{3}-b_{3}^{0}$ (where $b_{3}^{0}$ is the
free particle fugacity coefficient) in terms of sums and differences
of 3-body eigenphase shifts. Further, we show that if we obtain these
eigenphase shifts in a distorted-Born approximation, then, to first
order, we reproduce the leading low temperature behaviour, obtained
from an expansion of the two-fold integral of Dodd and Gibbs. The
contributions of the oscillating terms cancel.

The formalism that we propose is not limited to one dimension, but
seeks to provide a general method to obtain virial coefficients, fugacity
coefficients, in terms of asymptotic quantities. The exact one dimensional
results allow us to confirm the validity of our approach in this domain.


 
\end{abstract}


\pagebreak{}
\section*{Introduction}

As demonstrated in Nature\cite{nature} the behaviour of quantum models,
even in one dimension, and of such gases, remains of current interest!
In that paper, the paramount interest is the experimental investigation
of a system that is not ergotic and does not equilibrate after thousands
of collisions. Still, the description of a quantum mechanical gas,
in equilibrium, needs adequate description. So, for example, we note
the fairly recent article by Hussien and Yahya \cite{hussien}, on
the quantum corrections to the classical fourth virial coefficient
for square well potentials.

Our efforts are, instead, directed to the continuing development of
a fully quantum mechanical formalism, applicable to the higher virials,
and especially useful at low temperatures.

Let us be clear, there is an enormous difficulty in evaluating the
required traces in Statistical Mechanics, which involve 3 or more
particles. 
Our adiabatic hyperspherical formalism, uniquely, can `deal' with
3 or more quantum mechanical particles, whether for Boltzmann or Bose
or Fermi statistics, in arbitrary dimensions. This is very important:
it can be extended to more particles. The hyperspherical formalism
can, in principle, be used to represent an arbitrary number of particles.

Another `key' element is that we express our results in terms of `asymptotic'
quantities, in our particular case here: `eigenphase shifts'. This
can be done for the 2-body problem, and the essential `jump' is from
two to three and more particles!

We are continuing to extend our formalism, and we use 1-dim to validate
our results and to verify that we can carry out our fugacity program.
In this paper, we show how we `deal' with Bose statistics and repulsive
forces. Previously, we have limited ourselves to Boltzmann statistics,
but developed the present approach which relies on the use of a 
complete hyperspherical adiabatic basis\cite{larsen}. 

\newpage{}

We recall that, many years ago, Dodd and Gibbs \cite{dodd} were able
to evaluate the second and third virials of identical Bose particles
in one dimension, subject to repulsive delta function interactions.
To do this, they introduced a complete set of energy eigenfunctions
in the traces, and then integrated to obtain answers in terms of single
and two-dimensional integrals.

Their original aim, as is clear from their paper and from Gibbs' thesis,
was to obtain a formalism in terms of the S-matrix, following the
formalism of Dashen, Ma and Bernstein \cite{dashen}, in an effort
to obtain the virials in terms of scattering quantities. This was
not possible, as the authors determined that the {}``formal limiting
processes are not valid for the singular amplitudes of this system''.

An alternative formalism, proposed by Larsen and Mascheroni \cite{masche},
was a step aimed at ultimately expressing the higher virials in terms
of eigenphase shifts and bound state energies, and, if necessary,
other on-shell properties, in a hyperspherical harmonic formalism.
Under some constraints, one was then able to recover the classical
results for the 3rd virial, through a semi-classical approximation
\cite{larsen2}. Low temperature calculations have been carried out
for repulsive potentials in 2 dimensions \cite{jei}, and a more sophisticated
formalism, potentially able to `handle' correctly the bound states,
has been presented for Boltzmann statistics\cite{larsen}. It involves
the use, from the beginning, of a hyperspherical adiabatic basis.
Here, we extend that formalism to apply to Bose gases.

To summarize, our aim, in this paper, is to use the excellent results
of Dodd and Gibbs to confirm our virial formalism for the third fugacity
coefficient, for this particular dimension and interaction, for which
we have analytical and numerical results \cite{gibson,popiel,amaya}.
More specifically, we will show that the relevant $b_{3}-b_{3}^{0}$ can 
be rewritten as an integral, over the energy,
of a Boltzmann factor together with the sums and differences of eigenphase
shifts, for 3-body wave functions in one dimension. Using, then, a
distorted-Born approximation for the eigenphase shifts, we reproduce
the leading low temperature behaviour, obtained from an expansion
of the two-fold integral of Dodd and Gibbs. \newpage{}

\section*{Statistical Mechanics}

\noindent 
In this section we start with the Grand Partition function in 3 dimensions:

\begin{equation}
{\cal Q}={\displaystyle \sum_{n=0}^{\infty}z^{n}{}Tr\left(e^{-\beta H_{n}}\right)}=1+zTr\left(e^{-\beta T_{1}}\right)+z^{2}{}Tr\left(e^{-\beta H_{2}}\right)+z^{3}{}Tr\left(e^{-\beta H_{3}}\right)+\cdots\label{part}\end{equation}
 The fugacity $z$ equals $\exp\left(\mu/\kappa T\right)$;$\;\beta=1/\kappa T$,
where $\mu$, $\kappa$ and $T$ are the Gibbs' function per particle,
Boltzmann's constant and the temperature, respectively. $H_{n}$ and
$T_{n}$ are the n-particle Hamiltonian and kinetic energy operators.

We note that no factorials (1/factorials) appear in this development.
This is correct for Bose or Fermi statistics. 
Also important, is the fact that \[
Tr^{}\left(e^{-\beta H_{n}}\right)\rightarrow V^{n}\;\; as\; 
the \; volume \; \; V \; \rightarrow \; large,\]
 leading to the divergence of the individual traces in the thermodynamic
limit. If we take, however, the logarithm of the Grand Partition function
${\cal Q}$, we obtain \begin{equation}
\begin{array}{cl}
ln{\cal Q} & =z\,\, Tr\left(e^{-\beta T_{1}}\right)\\
 & +z^{2}\left[Tr\left(e^{-\beta H_{2}}\right)-\frac{1}{2}\left[Tr\left(e^{-\beta T_{1}}\right)\right]^{2}\right]\\
 & +z^{3}\left[Tr\left(e^{-\beta H_{3}}\right)-Tr\left(e^{-\beta T_{1}}\right)Tr\left(e^{-\beta H_{2}}\right)+\frac{1}{3}\left[Tr\left(e^{-\beta T_{1}}\right)\right]^{3}\right]\\
 & +\cdots\end{array}\label{lnq}\end{equation}
 which, when divided by V, gives coefficients of the powers of $z$,
which are independent of the volume, when the latter becomes large.
They are the fugacity coefficients $b_{l}$. \esp We can then write
for the pressure and the density \[
p/\kappa T=(1/V)\ln{\cal Q}=\sum_{l}b_{l}\, z^{l}\]
 \[
N/V=\rho=\sum_{l}l\, b_{l}\, z^{l},\]
 and the fugacity can be eliminated to yield the pressure in terms
of the density, in the virial expansion: \[
\frac{p}{\mathrm{k}T}=\sum_{n=1}^{\infty}B_{n}\rho^{n}.\]

\subsection*{One dimension, three Bose particles}

We follow the procedures outlined in the `Bogoliubov' lecture\cite{larsen},
developed for Boltzmann particles, and adapt the formalism to Bose
Statistics.

From Eq.(\ref{lnq}), and changing from the volume $V$ to a length
$L$, an expression for the 3rd fugacity coefficient can be written
as \begin{equation}
b_{3}=\frac{1}{L}\left[Tr\left(e^{-\beta H_{3}}\right)-Tr\left(e^{-\beta T_{1}}\right)Tr\left(e^{-\beta H_{2}}\right)+\frac{1}{3}Tr\left(e^{-\beta T_{1}}\right)^{3}\right]\label{b3}\end{equation}
 Note that each of the 3 terms in the bracket, above, grows as $L^{3}$
asymptotically, for large $L$. Therefore, subtractions must take
place, so that $b_{3}$ might become $L$-independent, as is required.
If we factor out the contribution of the C.M. (proportional to $L$),
each term will have a dependence proportional to $L^{2}$.

To help us decrease the $L$ dependence, let us subtract, from each
of terms above, the equivalent term without interaction.

Subtracting, therefore, $b_{3}^{0}$ from the $b_{3}$, we obtain:
\begin{equation}
b_{3}-b_{3}^{0}=\frac{1}{L}\left[Tr\left(e^{-\beta H_{3}}-e^{-\beta T_{3}}\right)-Tr\left(e^{-\beta T_{1}}\right)Tr\left(e^{-\beta H_{2}}-e^{-\beta T_{2}}\right)\right]\label{b30a}\end{equation}
 \newpage{}

\section*{Adiabatic Preliminaries}

For the 3 particles of equal masses, in one dimension, we first introduce
center of mass and Jacobi coordinates. We define \[
{\eta}=\left(\frac{1}{2}\right)^{1/2}({x}_{1}-{x}_{2})\hsp{\xi}=\left(\frac{2}{3}\right)^{1/2}\left(\frac{{x}_{1}+{x}_{2}}{2}-{x}_{3}\right)\hsp{R}=\,\;\frac{1}{3}\;({x}_{1}+{x}_{2}+{x}_{3})\]
 where, of course, the ${x}_{i}$ give us the locations of the 3 particles.
This is a canonical transformation and insures that in the kinetic
energy there are no cross terms.

The variables ${\xi}$ and ${\eta}$ are involved separately in the
Laplacians and we may consider them as acting in different spaces.
We introduce a higher dimensional vector $\vec{\rho}=\left(\begin{array}{c}
{\xi}\\
{\eta}\end{array}\right)$ and express it, here in two dimensions, in a simple polar coordinate
system (using the radius $\rho$ and the angle $\vartheta$). 
We write \begin{equation}
\eta=\rho\cos{\vartheta};\hspace{.2in}\xi=\rho\sin{\vartheta},\hspace{.2in}-\pi<\vartheta\leq\pi;\hspace{.2in}0\leq\rho<\infty.\end{equation}

If we factor a term of $\rho^{1/2}$ from the solution of the relative
Schr\"odinger equation, i.e. we let $\psi=\tilde{\phi}/\rho^{1/2}$,
we are led to: \[
\left[-\frac{\partial^{2}}{\partial\rho^{2}}+H_{\rho}-k^{2}\right]\tilde{\phi}(k,\overrightarrow\rho)=0\]
 where \[
H_{\rho}=-\frac{1}{\rho^{2}}\left[\frac{\partial^{2}}{\partial\vartheta^{2}}+\frac{1}{4}\right]+\frac{2m}{\hbar^{2}}V(\rho,\vartheta),\]
$m$ is the mass of each particle, $k=\sqrt{2mE/\hbar^{2}}$ and $E$ is the relative energy in the center of mass. \\
 We now introduce the adiabatic basis, which consists of the eigenfunctions
of part of the Hamiltonian: the angular part of the kinetic energy
and the potential. \begin{equation}
H_{\rho}B_{\ell}(\rho,\vartheta)=\Lambda_{\ell}(\rho)B_{\ell}(\rho,\vartheta),\label{ang}\end{equation}
 where $\ell$ enumerates the solutions.

Using this adiabatic basis, we can now rewrite the Schr\"{o}dinger equation
as a system of coupled ordinary differential equations. For a typical
solution, we write \begin{equation}
\tilde{\phi}(k,\overrightarrow\rho)=\sum_{\ell^{\prime}}B_{\ell^{\prime}}(\rho,\vartheta)\,\phi_{\ell^{\prime}}(k,\rho)\end{equation}
 and obtain the set of coupled equations \begin{eqnarray}
\left(\frac{d^{2}}{d\rho^{2}}-\Lambda_{\ell}(\rho)  +   k^{2}\right)\,\phi_{\ell}(k,\rho) &+ & 2\sum_{\ell^{\prime}}C_{\ell}^{\ell^{\prime}}\left(\rho\right)\,\frac{d}{d\rho}\phi_{\ell^{\prime}}(k,\rho)\nonumber \\
 & + & \sum_{\ell^{\prime}}D_{\ell}^{\ell^{\prime}}\left(\rho\right)\,\phi_{\ell^{\prime}}(k,\rho)=0,\label{diff}\end{eqnarray}
where we defined: \begin{eqnarray}
C_{\ell}^{\ell^{\prime}}(\rho) & = & \int d\vartheta\, B_{\ell}^{\ast}(\vartheta,\rho)\frac{\partial}{\partial\rho}B_{\ell^{\prime}}(\vartheta,\rho)\label{Cs}\\
D_{\ell}^{\ell^{\prime}}(\rho) & = & \int d\vartheta\, B_{\ell}^{\ast}(\vartheta,\rho)\frac{\partial^{2}}{\partial\rho^{2}}B_{\ell^{\prime}}(\vartheta,\rho).\label{Ds}\end{eqnarray}

\subsection*{Three binary interactions: case  $\left(123\right)$}

\noindent In our model calculation, the interaction between our particles
will be a repulsive delta-function, with `strength' $g>0$. For 3
particles, we will thus have an interaction potential $V(\rho,\vartheta)$:
\begin{equation}
V(\rho,\vartheta)=g[\delta(\sqrt{2}\rho|\cos\vartheta|)+\delta(\sqrt{2}\rho|\cos(\vartheta-2\pi/3)|)+\delta(\sqrt{2}\rho|\cos(\vartheta+2\pi/3)|)].\end{equation}
 In a set of previous papers [9-15]
we, and colleagues, have studied these systems (with $g$
positive and negative), calculated eigenpotentials, adiabatic bases,
phase shifts, eigenfunctions and binding energies - where appropriate. 
We refer to the Appendices for a selection of results, that we need in 
our present calculations.

\esp The adiabatic eigenfunctions $B_{K}(\rho,\vartheta)$ can be
characterized by their symmetry, together with the index $K$, as
they reduce to simple normalized surface harmonics when the potential
$V(\rho,\vartheta)\rightarrow0$ and their associated eigenvalues
$\Lambda_{K}\rightarrow(K^{2}-\frac{1}{4})/\rho^{2}$.

\esp For these surface harmonics, in the completely symmetric 1-dimensional
representation $\Gamma_{1}$, we need \cite{larsen0} to set $K=0\, mod\,3$
. The harmonics will then be:\begin{equation}
\begin{array}{cl}
\mathrm{\textrm{for}}\: K=\textrm{even}, & \frac{1}{\sqrt{\pi}}\cos(K\vartheta)\qquad\textrm{with}\;\frac{1}{\sqrt{2\pi}}\;\textrm{for}\; K=0\\
\mathrm{\textrm{for}}\: K=\textrm{odd}, & \frac{1}{\sqrt{\pi}}\sin(K\vartheta)\end{array}\label{eq:har123}\end{equation}

\esp {[}The completely antisymmetric 1-dimensional representation
$\overline{\Gamma}_{1}$ requires, in the above, to interchange the
role of the sines and cosines (clearly without the constant term).
Finally, in cases where $K\neq0\, mod\,3$, the sines and cosines
form pairs belonging to a 2-dimensional representation $\Gamma_{2}$.
] \\

\noindent The $B_{K}(\rho,\vartheta)$ can then be constructed from
the harmonics or obtained as solutions of a differential equation,
Eq.(\ref{ang}). The resulting forms are then\cite{vinit}: \\
\begin{equation}
\begin{array}{ccl}
\textrm{for}\: K=\textrm{even}, & A_{K}(\rho)\,\cos(q_{K}(\vartheta-\frac{m\pi}{3})) & \textrm{where}\: q_{K}\:\textrm{satisfies}\quad q_{K}\tan(\frac{\pi q_{K}}{6})=\frac{\pi\rho c}{6}\\
\mathrm{\textrm{for}}\: K=\textrm{odd}, & A_{K}(\rho)\,\sin(q_{K}(\vartheta-\frac{m\pi}{3})) & \textrm{where}\: q_{K}\:\textrm{satisfies}\quad q_{K}\cot(\frac{\pi q_{K}}{6})=-\frac{\pi\rho c}{6}\end{array}\label{eq:ad123}\end{equation}

\noindent and where $m$ is an integer such that $|\vartheta-m\pi/3|<\pi/6$
and $c=(2m/\hbar^{2})(3g/\pi\sqrt{2})$. \\
 For each value of $\rho$, we need to solve the transcendental equation
for $q_{K}$. We can do so numerically, or alternatively seek expansions
in $\rho$ or $1/\rho$, for small or large values of $\rho$, respectively.
\\
 $A_{K}(\rho)$ is simply the normalizing factor, such that the $B_{K}(\rho,\vartheta)$
are orthonormal when integrating over the angle, with a choice of
integrating over a `sector', say $-\pi/6$ to $\pi/6$, or over a
more extended range such as from $0$ to $2\pi$.

\subsection*{Two interacting particles + a spectator: case  $\left(12, 3 
\right)$}

In this case, which we also require, we use our 3-body formalism to
treat three particles, only two of which are interacting, the other
one playing the role of a spectator. It is very important that this
be done in a way closely paralleling the treatment for the three interacting
particles, because we will need to subtract traces, associated with
lengths (volumes in 3-dimensions) and they must be evaluated in a
similar way in all of the terms of our expressions.

Since our need is the evaluation of traces, we can choose particles
$1$ and $2$ as interacting, particle $3$ as the spectator. Letting
$\vartheta\rightarrow-\vartheta+\pi$, and therefore letting $\eta$
go to $-\eta$ and $\xi$ go to $\xi$, interchanges the role of particles
$1$ and $2$. The symmetrical combinations for the harmonics (i.e.
which treat particles $1$ and $2$ in a symmetrical fashion) will
then be: \\
\begin{equation}
\begin{array}{cl}
\textrm{for}\: K=\textrm{even}, & \frac{1}{\sqrt{\pi}}\cos(K\vartheta)\qquad\textrm{with}\frac{1}{\sqrt{2\pi}}\;\textrm{for}\; K=0\\
\mathrm{\textrm{for}}\: K=\textrm{odd}, & \frac{1}{\sqrt{\pi}}\sin(K\vartheta)
\end{array}
\label{eq:har12,3}
\end{equation}
\esp where, of course, there is no longer a restriction of $0\, mod\,3$.
$K$ now runs over $0,2,4,6...$ or $1,3,5...$.

\noindent \newpage{} 
The $B_{K}^{12,3}(\rho,\vartheta)$ will read: \\
\begin{equation}
\begin{array}{ccc}
\textrm{for}\: K=\textrm{even}, & 
A_{K}^{12,3}(\rho)\,\cos(q_{K}(\vartheta-m\pi)) & 
\textrm{where}\: 
q_{K}\:\textrm{satisfies}
\quad q_{K}\tan(\frac{\pi q_{K}}{2})=\frac{\pi\rho c}{6}\\ \,\\
\mathrm{\textrm{for}}\: K=\textrm{odd}, & 
A_{K}^{12,3}(\rho)\,\sin(q_{K}(\vartheta-m\pi)) & \; \;  \, 
\textrm{where}\: 
q_{K} \: \textrm{satisfies}
\quad q_{K}\cot(\frac{\pi q_{K}}{2})=-\frac{\pi\rho c}{6}
\end{array}
\label{eq:ad12,3}
\end{equation}

\noindent and where $m$ is an integer such that $|\vartheta-m\pi|<\pi/2$
and, again, $c=(2m/\hbar^{2})(3g/\pi\sqrt{2})$. \\

The $A_{K}^{12,3}(\rho)$ are the normalizing factors, such that the
$B_{K}^{12,3}(\rho,\vartheta)$ are orthonormal, when integrated over
the angle - with a choice of integrating over a `sector' (here the
right-hand or left-hand part of the range) or over the whole range of
$0$ to $2\pi$. \\
 The $q_{K}$, here, are really $q_{K}^{12,3}$ and, of course, as
in the fully interacting case, are also functions of $\rho$.

\subsection*{$\Lambda$'s, $C$'s, $D$'s}

Given the expressions for the $B$'s ... we can then obtain expansions
for the behaviour of the $q_{K}$ as functions of $\rho$, and therefore
for $\Lambda_{K}$ and for the $C$'s and the $D$'s .... For us,
the most important behaviour is that as $\rho$ becomes large ....
i.e. expansions in $1/\rho$.

\noindent We then find, {[}see Appendix B], that for $\rho\rightarrow\infty$
the \textit{effective} centrifugal term (i.e. the $1/\rho^{2}$ term)
reads :\begin{equation}
\begin{array}{ccc}
\Lambda_{K} & =
\frac{(K+3)^{2}}{\rho^{2}}\;+\;\ldots & \textrm{for the case} \; 
\left(123\right) \\
\,\\
\Lambda_{K}^{12,3} & =\frac{(K+1)^{2}}{\rho^{2}}\;+\;\ldots & \; 
\textrm{for the case} 
\;
\left(12,3\right)
\end{array}
\label{eq:int1}
\end{equation}
and in both cases, the $C$ and $D$ terms behave as : \begin{equation}
\begin{array}{rcl}
D{}_{K}^{K'} & \propto & \left\{ \begin{array}{ll}
\frac{1}{\rho^{3}}+\ldots & K\not=K'\\
\frac{1}{\rho^{4}}+\ldots & K=K'\end{array}\right.\\
\; & \; & \;\\
C_{K}^{K'}\;\frac{d}{d\rho} & \propto & \left\{ \begin{array}{ll}
\frac{1}{\rho^{2}}\;\frac{d}{d\rho}+\ldots & K\not=K'\\
\; & \;\\
0 & K=K'\end{array}\right.\end{array}\label{int3}\end{equation}
In both cases, to obtain a non-zero result, the indices of the $K$
and $K^{\prime}$ matrix elements must both belong either to an order
which is even or to an order which is odd. There are no cross terms.

\subsection*{Coupled equations and asymptotic behaviour}

Given N coupled equations, we will have N independent solutions (well
behaved at the origin) of the coupled differential equations. These
can be expressed as: \begin{equation}
\tilde{\phi}^{K}(k,\overrightarrow\rho)=\sum_{K^{\prime}}B_{K^{\prime}}(\rho,\vartheta)\,\phi_{K^{\prime}}^{K}(k,\rho)\label{eq:fi}\end{equation}
 where the $\phi_{K'}^{K}$ - the amplitudes - can be chosen to have
as asymptotic behaviour, either: \begin{equation}
\phi_{K'}^{K}(k,\rho)=\delta_{K'}^{K}\;\sqrt{k\rho}\; J_{K^{\prime}+3}(k\rho)\;+\; W_{K'}^{K}(k)\;\sqrt{k\rho}\; N_{K^{\prime}+3}(k\rho)\label{eq:fi123}\end{equation}
 or \begin{equation}
\bar{\phi}_{K'}^{K}(k,\rho)=\delta_{K'}^{K}\;\sqrt{k\rho}\; J_{K^{\prime}+1}(k\rho)\;+\;\overline{W}_{K'}^{K}(k)\;\sqrt{k\rho}\; N_{K^{\prime}+1}(k\rho),\label{eq:fi12,3}\end{equation}
 as appropriate for the desired case. \\

\noindent 
[I.e. starting from amplitudes with $A_{K^{\prime}}^{K}$
and $B_{K^{\prime}}^{K}$ as coefficients of the Bessel and Neumann
functions (multiplied by $\sqrt{k\rho}$), we take linear combinations
of the solutions - multiplying by the matrix $A^{-1}$ - to obtain
equations (\ref{eq:fi123}) and (\ref{eq:fi12,3}).] The upper index
$K$, chosen to enumerate the solutions (\ref{eq:fi}), is associated
with the $J_{K}(k\rho)$ which dominates the solution for small values
of $\rho$, and takes on the same set of values as the lower index
$K^{\prime}$ . \\
 The wave functions and the amplitudes can be chosen to be real, and
the $W_{K}^{K'}$ and the $\overline{W}_{K}^{K'}$ will be real and
symmetrical. They can be diagonalized to yield tangents of eigenphase
shifts $-\tan{\delta_{W}^{K}}(k)$ and $-\tan{((\bar{\delta}_{W})^{K})}(k)$,
respectively. \\

Further, we can then take linear combinations of these solutions,
denoted as $\phi_{K^{\prime}}^{\nu}$, such that for the fully interacting
system : \begin{equation}
\phi_{K^{\prime}}^{\nu}(k,\rho)\rightarrow(k\rho)^{1/2}\mathcal{\mathcal{\mathcal{C}}}_{K^{\prime}}^{\nu}(k)\,[\cos\delta_{W}^{\nu}(k)\: J_{K^{\prime}+3}(k\rho)-\sin\delta_{W}^{\nu}(k)\: N_{K^{\prime}+3}(k\rho)]\label{eq:finu}\end{equation}
 where $K^{\prime}$ denotes the $K^{\prime}$th amplitude of the
solution $\nu$ and the eigenphase shifts are associated with the
matrix $W$. In the fully interacting case, these phase shifts will
only differ by $-3\pi/2$ from those of the genuine $S$-matrix for
this problem. (I.e. $-3\pi/2+\delta_{W}=\delta_{S}$) \\
 In the (12,3) case, we obtain similar expressions. \\

\noindent We note that since the \, `$K^{\prime}$ = even' \, amplitudes
couple only with the \, `$K^{\prime}$ = even' \, amplitudes and,
analogously, the \, `$K^{\prime}$ = odd' \, amplitudes couple only
with the \, `$K^{\prime}$ = odd' \, amplitudes, we can therefore
separate our set of solutions $\phi_{K^{\prime}}^{\nu}$ into 2 sets,
one with amplitudes involving only even values of $K^{\prime}$ and
the other with amplitudes involving only odd values of $K^{\prime}$.

\newpage{}

\section*{The phase shift expressions for the Traces}

\subsection*{1st term - with the totally symmetric adiabatic $\left(123\right)$
basis}

Consider the first term of the Eq.$\left(\ref{b30a}\right)$ :

\begin{equation}
\frac{3^{1/2}}{\lambda_{T}}\left[Tr\left(e^{-\beta H_{3}^{rel}}\right)-Tr\left(e^{-\beta T_{3}^{rel}}\right)\right],\label{tr3}\end{equation}

\noindent seen here, after we have separated the Hamiltonian into
two parts, one corresponding to the center of mass coordinate and
the other to the relative motion, and taken the trace of the center
of mass contribution. The latter contributes a factor that depends
on the temperature $\lambda_{T}=(h^{2}/2\pi m\kappa T)^{1/2}$ and
a length that cancels the factor $1/L$ shown in Eq.$\left(\ref{b30a}\right)$.
The first term of the above can now be written as

\begin{equation}
\frac{3^{1/2}}{\lambda_{T}}\int d\overrightarrow{\rho}\int_{0}^{\infty}dk\,\sum_{\nu}\psi^{\nu*}\left(k,\overrightarrow{\rho}\right)\psi^{\nu}\left(k,\overrightarrow{\rho}\right)e^{-\beta\left(\hbar^{2}/2m\right)k^{2}}\label{tr3a}\end{equation}
where we introduced a complete set of energy
eigenfunctions, defined over the hyperspherical space $\overrightarrow{\rho}$.
\\

As we have indicated before, we let $\psi^{\nu}=\tilde{\phi}^{\nu}/\rho^{1/2}$,
and then expand the eigenfunction $\tilde{\phi}^{\nu}$
in terms of an adiabatic basis $B_{K}^{123}$,
which is completely symmetric, complete and normalized over each of
the intervals $\left(m\pi/3-\pi/6,m\pi/3+\pi/6\right)$, where $m=0,1,\cdots5$.
Each of these intervals is associated with one of the 6 possible permutations
of the coordinates of the system. Integrating over just one of the
sectors is to integrate over all of the possible configurations of
the physical particles in a Bose system.

Alternatively, if we wish to integrate over the interval $-\pi$ to
$\pi$ we can divide the angular functions by $\sqrt{3!}$, taking
into account their six-fold symmetry, and obtain a symmetric adiabatic
base $B_{K}\left(\rho,\vartheta\right)$, normalized over this extended
angular interval. Expanding the functions in terms of the symmetric
adiabatic base, expression $\left(\ref{tr3a}\right)$ now reads:

\begin{equation}
\begin{array}{c}
\frac{3^{1/2}}{\lambda_{T}}\int_{0}^{\infty}dk\int_{0}^{\infty}d\rho{\displaystyle \sum_{\nu,K,K'}}\phi_{K}^{\nu*}\left(k,\rho\right)\phi_{K'}^{\nu}\left(k,\rho\right)\\
\int_{-\pi}^{\pi}d\vartheta B_{K}^{*}\left(\rho,\vartheta\right)B_{K'}\left(\rho,\vartheta\right)e^{-\beta\left(\hbar^{2}/2m\right)k^{2}}\end{array}\label{tr3b}\end{equation}
 Here $\nu$ stands for the $\nu$th solution and $K$, which takes
on only even or only odd values for a given $\nu$, labels the components.
Integration over $\vartheta$ then gives

\begin{equation}
\frac{3^{1/2}}{\lambda_{T}}\int_{0}^{\infty}dk\sum_{\nu,K}\int_{0}^{\infty}d\rho\,\,\phi_{K}^{\nu*}\left(k,\rho\right)\phi_{K}^{\nu}\left(k,\rho\right)e^{-\beta\left(\hbar^{2}/2m\right)k^{2}}\label{tr3c}\end{equation}

We note that had we expanded in terms of $B_{K}^{123}\left(\rho,\vartheta\right)$,
and integrated over one of the sectors, we would have obtained the
same expression.

The radial functions satisfy the coupled radial equations: \begin{eqnarray}
\left(\frac{d^{2}}{d\rho^{2}}-\Lambda_{K}(\rho)  +  k^{2}\right)\,{\phi}_{K}^{\nu}(k,\rho) &+ & 2\sum_{K^{\prime}}C_{K}^{K^{\prime}}(\rho)\,\frac{d}{d\rho}{\phi}_{K^{\prime}}^{\nu}(k,\rho)\nonumber \\
 & + & \sum_{K^{\prime}}D_{K}^{K^{\prime}}(\rho)\,{\phi}_{K^{\prime}}^{\nu}(k,\rho)=0\end{eqnarray}
and we can choose our solutions to be real.

To evaluate the integral of the square of the wave function over $\rho$
we now use a procedure similar to that used in our previous papers\cite{2nd}.
We integrate the square of the wave functions within a sphere of radius
$\rho_{max}$, and then let $\rho_{max}$ go to infinity. We use the
Green's function `trick'\cite{larsen}, and obtain

\begin{eqnarray}
\int_{0}^{\rho_{max}}\!\sum_{K}\;(\!\! & \!\phi_{K}^{\nu}(k,\rho) & \!\!\phi_{K}^{\nu}(k^{\prime},\rho))d\rho=\nonumber \\
\frac{1}{k^{2}-(k^{\prime})^{2}}\sum_{K}\;[\! & \!\!\phi_{K}^{\nu}(k,\rho) & \!\!\frac{d}{d\rho}\phi_{K}^{\nu}(k^{\prime},\rho)-\phi_{K}^{\nu}(k^{\prime},\rho)\frac{d}{d\rho}\phi_{K}^{\nu}(k,\rho)],\label{green}\end{eqnarray}
 evaluated at $\rho=\rho_{max}$. \\
 ----------------------------------------------------------------------
\\
 I.e. our identity is:\begin{equation}
\begin{array}{ll}
{\displaystyle \;\sum_{K}}\frac{d}{d\rho}\left[\phi_{K}^{\nu}(k^{\prime},\rho)\frac{d}{d\rho}\phi_{K}^{\nu}(k,\rho)-\phi_{K}^{\nu}(k,\rho)\frac{d}{d\rho}\phi_{K}^{\nu}(k^{\prime},\rho)\right]\\
\,\\
+\left(k^{2}-(k^{\prime})^{2}\right){\displaystyle \sum_{K}}\phi_{K}^{\nu}(k,\rho)\,\phi_{K}^{\nu}(k^{\prime},\rho)\\
+\:2\ {\displaystyle \sum_{K,{K}^{\prime}}}\frac{d}{d\rho}\left[\phi_{K}^{\nu}(k^{\prime},\rho)\ C_{K}^{K^{\prime}}\left(\rho\right)\ \phi_{{K}^{\prime}}^{\nu}(k,\rho)\right] & =0\end{array}\label{eq:trick}\end{equation}
and we integrate with respect to $\rho$. Using then the fact that
$\phi_{K}^{\nu}$ goes to zero, as $\rho$ itself goes to zero, and that $C_{K}^{K^{\prime}}$ decreases
fast enough for $\rho$ large, we are left with the expression displayed
earlier (that of our `trick'). \\
 ---------------------------------------------------------------------
\\

\noindent We now put in the real version of the asymptotic form of our 
solutions, Eq.(\ref{eq:finu}), oscillatory solutions valid for $\rho_{max}$
large, and use l'Hospital's rule to take the limit as $k^{\prime}\rightarrow k$.
Thus for the interacting system 
\begin{equation}
\phi_{K}^{\nu}(k,\rho) \rightarrow(k\rho)^{1/2}
\mathcal{C}_{K}^{\nu}(k)\,[\cos\delta_{W}^{\nu}(k)\: J_{K+3}
(k\rho)-\sin\delta_{W}^{\nu}(k)\: N_{K+3}(k\rho)].
\label{beauty}\end{equation}
 Here, we note again that we come to a major difference with what
happens in other dimensions. We have seen, Eqs. (\ref{eq:int1}), that
for large values of $\rho$, the effective asymptotic `centrifugal'
term involves $K+3$, instead of the $K$ that appears for small values
of $\rho$. The corresponding `free' particle solution can be chosen
to be: \[
(\phi^{0})_{K}^{\nu}(k,\rho)=(k\rho)^{1/2}\mathcal{C}_{K}^{\nu}(k)\, J_{K}(k\rho)\]

\noindent We note that $\delta^{\nu}$ is associated with the solution
$\nu$, and appears in each of the amplitudes of this particular solution.
The coefficient $\mathcal{C}_{K}^{\nu}$, the mixture coefficient,
tells us how much of each normalized amplitude appears in the solution.
These two quantities therefore completely characterize the asymptotic
behaviour of the $\nu$th wave function. We expect that \[
{\displaystyle \sum_{K}(\mathcal{C}_{K}^{\nu})^{*}}\mathcal{C}_{K}^{\nu}=1.\]
 (Or the real equivalent.)

Inserting this into our integrals, and taking the limit 
$k^{\prime}\rightarrow k$, we find that \[
\sum_{K}\int_{0}^{\rho_{max}}\!|\phi_{K}^{\nu}(k,\rho)|^{2}\: d\rho\rightarrow\frac{1}{\pi}\frac{d}{dk}\delta^{\nu}(k)+\frac{1}{\pi}\rho_{max}\:+\: osc.\: terms\]
 and, thus, that \begin{equation}
\sum_{K}\int_{0}^{\rho_{max}}\!(|\phi_{K}^{\nu}(k,\rho)|^{2}-|(\phi^{0})_{K}^{\nu}(k,\rho)|^{2})\; d\rho\rightarrow\frac{1}{\pi}\frac{d}{dk}\delta^{\nu}(k)\:+\: osc.\: terms\label{ddelta}\end{equation}
 We let $\rho_{max}$ go to infinity, and usually the oscillating
terms do not contribute to the integral over the energy. In 1 dimension,
however, we have to be more careful.

\newpage{}
\noindent
Using the asymptotic form of real amplitudes

\begin{equation}
\sqrt{\frac{2}{\pi}}\mathcal{C}_{K}^{\nu}(k)\sin\left[k\rho-\frac{1}{2}(K+3)\pi+\frac{1}{4}\pi+\delta^{\nu}\left(k\right)\right]\label{fia3}\end{equation}
 and l'Hospital's rule, when taking the limit as $k'\rightarrow k$
in Eq.$\left(\ref{green}\right)$, we find that the integral over
$\rho$ , can be written as

\begin{equation}
\begin{array}{l}
\frac{1}{\pi}{\displaystyle \sum_{\nu}}\left[\rho_{max}+\frac{d}{dk}\delta^{\nu}\left(k\right)\right]+\frac{1}{2\pi}{\displaystyle \sum_{\nu,K}}\left[\mathcal{C}_{K}^{\nu}(k)\right]^{2}\frac{1}{k}\sin\left[2\left(k\rho_{max}-\frac{1}{2}(K+3)\pi+\frac{1}{4}\pi+\delta^{\nu}\left(k\right)\right)\right]\end{array}\label{eq:a}\end{equation}
 When we {\bf subtract} the term without interaction from the above expression,
the integral over k takes the form, involving the eigenphases:

\begin{equation}
\begin{array}{l}
\frac{3^{1/2}}{\lambda_{T}}\int_{0}^{\infty}dk\, e^{-\beta\left(\hbar^{2}/2m\right)k^{2}}\,\frac{1}{\pi}\,{\displaystyle \sum_{\nu}}\left[\frac{d}{dk}\delta^{\nu}(k)\right]
\end{array}
\label{eq:b}
\end{equation}
 and the contribution from the oscillating terms: \begin{equation}
\begin{array}{l}
\frac{3^{1/2}}{\lambda_{T}}\int_{0}^{\infty}dk\frac{1}{2\pi}{\displaystyle \sum_{\nu,K}}(\mathcal{C}_{K}^{\nu}(k))^{2}\left\{ \frac{1}{k}\left(\sin\left[2\left(k\rho_{max}-\frac{1}{2}(K+3)\pi+\frac{1}{4}\pi+\delta^{\nu}\left(k\right)\right)\right]\right.\right.\\
\left.\left.-\sin\left[2\left(k\rho_{max}-\frac{1}{2}K\pi+\frac{1}{4}\pi\right)\right]\right)\right\} e^{-\beta\left(\hbar^{2}/2m\right)k^{2}}\end{array}\label{eq:b2}\end{equation}
 which can be rewritten as \begin{equation}
\begin{array}{c}
\frac{3^{1/2}}{\pi\lambda_{T}}\int_{0}^{\infty}dk{\displaystyle \sum_{\nu}}\,(\mp)_{\nu}\,\left\{ \frac{1}{2k}\left(\cos\left[2\left(k\rho_{max}\right)\right]\left[\cos\left(2\delta^{\nu}\left(k\right)\right)+1\right]\right.\right.\\
\left.\left.-\sin\left[2\left(k\rho_{max}\right)\right]\sin\left(2\delta^{\nu}\left(k\right)\right)\right)\right\} e^{-\beta\left(\hbar^{2}/2m\right)k^{2}}\end{array}\label{eq:c}\end{equation}
 where we have used the fact that the $K$ are either all even or
all odd for a given eigenfunction $\nu$, giving a value of $-1$
to $\mp$, when the K are even and $+1$ when they are odd. For a
given $\nu$, we can therefore sum over $K$. In fact, since we have
isolated the $K$ dependence of the trigonometric terms, we can sum
the $(\mathcal{C}_{K}^{\nu})^{2}$, over $K$.

\noindent We should also mention that when we perform the subtraction
of the `free' wave function from the interacting one ... we must choose
precisely the same $\mathcal{C}'s$ for the free solution, as the
ones that are mandated when we have an interaction! We are however
free to do so!

The phase shifts, that appear in the expressions above, are the $W$-phase
shifts ... that go to zero as $k$ tends to zero. In our last equation,
therefore, the leading term of $cos(2\delta)+1$ will give $+2$ ...
and the individual integrands will `blow' at the origin. We have a
major problem! Unless our reasoning is incorrect, our only hope is
that alternating signs and other similar oscillatory terms - arising
from the $(12,3)$ contributions - will enable us to precisely cancel
what we have here. The term $\sin\left[2\left(k\rho_{max}\right)\right]\sin\left(2\delta^{\nu}\left(k\right)\right)\,/{k}$
will not contribute to the $k$-integral, when we take the limit $\rho_{max}\rightarrow\infty$
\vspace{0.2in}

\noindent \newpage{}

\subsection*{2nd term - with the symmetric adiabatic (12,3) basis}

We take the second set of traces in Eq.$\left(\ref{b3}\right)$ and
rewrite it as an expression involving a trace of a 3-body operator:
\begin{equation}
Tr\left(e^{-\beta T_{1}}\right)Tr\left(e^{-\beta H_{2}}\right)=Tr\left(e^{-\beta\left(H_{2}+T_{1}\right)}\right)\end{equation}
This is an essential trick. To obtain the L-independence of $b_{3}$,
we need to subtract related quantities from all of the traces, i.e.
evaluate them with same or similar harmonics or adiabatic functions,
have sums which, though they might individually diverge
if summed to infinity, can be subtracted to yield finite results.
They can each be evaluated for a finite number of
terms, subtracted by similar sums, and then permitted, for a fixed
energy, to converge [This has worked for a Boltzmann calculation\cite{jei}]
or - in our present case - subjected to an accelerated convergence procedure,
subtracted and found to yield a finite result. 

We now expand the wave functions, eigenstates of a 3-body Hamiltonian
of 3 particles, for which, say, particles 1 and 2 are interacting,
with particle 3 acting as a spectator, in terms of adiabatic angular
functions and $\rho$-dependent amplitudes.

\begin{equation}
\bar{\tilde{\phi}}^{\nu}\left(k,\overrightarrow\rho\right)=\frac{1}{\rho^{1/2}}\sum_{K}B_{K}^{12,3}\left(\rho,\vartheta\right)\overline{\phi}_{K}^{\nu}\left(k,\rho\right)\end{equation}
 [We note that $\nu$ and $K$ are not the same indices as appear
in the previous section, and do not assume the same sets of values.]
\\
 The angular functions, $B^{12,3}$, will be symmetric upon interchange
of particles 1 and 2, but will have no symmetry with respect to interchanges
between particles 1 and 3 or 2 and 3. They will be eigenfunctions
of Eq.$\left(\ref{ang}\right)$, with the potential $V$ equal to:
\[
\bar{V}=g\delta\left(|x_{1}-x_{2}|\right)=g\delta\left(\sqrt{2}\rho|\cos\vartheta|\right),\]
 and normalized over $1/2$ of the $\vartheta$ space, for example from
$-\pi/2$ to $+\pi/2$, since this corresponds to a complete sampling
of the possible physical configurations. Alternatively, we can define
a $\overline{B_{l}}$ equal to $B^{12,3}$ divided by $\sqrt{2}$,
which can be integrated from $-\pi$ to $+\pi$ and still be normalized.
The equivalent to Eq.$\left(\ref{tr3b}\right)$ is then \begin{equation}
\frac{3^{1/2}}{\lambda_{T}}\int_{0}^{\infty}dke^{-\beta\left(\hbar^{2}/2m\right)k^{2}}\int_{0}^{\infty}d\rho\sum_{\nu,K,K'}\left[\overline{\phi}_{K}^{\nu*}\left(k,\rho\right)\overline{\phi}_{K'}^{\nu}\left(k,\rho\right)\int_{-\pi}^{\pi}d\vartheta\,\overline{B}_{K}^{*}\left(\rho,\vartheta\right)\overline{B}_{K\prime}\left(\rho,\vartheta\right)\right]\end{equation}
 and the equivalent of Eq.$\left(\ref{tr3c}\right)$ is then \begin{equation}
\frac{3^{1/2}}{\lambda_{T}}\int_{0}^{\infty}dk\, e^{-\beta\left(\hbar^{2}/2m\right)k^{2}}\sum_{\nu,K}\int_{0}^{\infty}d\rho\,\overline{\phi}_{K}^{\nu*}\left(k,\rho\right)\overline{\phi}_{K}^{\nu}\left(k,\rho\right)\end{equation}

We can now subtract a trace with `free' particles, with wave functions
that are only symmetric under an interchange of particles 1 and 2,
integrate over $\rho$ and define eigenphase shifts $\overline{\delta}^{\nu}(k)$
for this 3-body system, following steps similar to those of the previous
section. There are differences, though. Instead of the $K+3$, which
appeared in the asymptotic behaviour of $\Lambda_{K}$, we find that
the effective asymptotic `centrifugal' term involves $K+1$. The phase
shifts must therefore be defined in terms of Bessel and Neumann functions
of order $K+1$. They are, again, $W$-type phase shifts ... which
go to zero as $k$ goes to zero.

For the contribution of the 2nd term, we then arrive at: 
\begin{equation}
\begin{array}{l}
\frac{3^{1/2}}{\lambda_{T}}\int_{0}^{\infty}dk\, 
e^{-\beta\left(\hbar^{2}/2m\right)k^{2}}\,\frac{1}{\pi}\,
{\displaystyle \sum_{\nu}}\left[\frac{d}{dk}
\overline{\delta}^{\nu}(k)\right]
\end{array}
\label{eq:b5}
\end{equation}
 where the sum over $\nu$ must involve the even and odd values of
$K$, but a particular $\overline{\psi}^{\nu}$ will only involve the even or
the odd values of $K$. The oscillating terms yield: 
\begin{equation}
\begin{array}{l}
\frac{3^{1/2}}{\lambda_{T}}\int_{0}^{\infty}dk\frac{1}{2\pi}{\displaystyle \sum_{\nu,K}}\left[\overline{\mathcal{C}}_{K}^{\nu}(k)\right]^{2}\left\{ \frac{1}{k}\left(\sin\left[2\left(k\rho_{max}-\frac{1}{2}(K+1)\pi+\frac{1}{4}\pi+\overline{\delta}^{\nu}\left(k\right)\right)\right]\right.\right.\\
\left.\left.-\sin\left[2\left(k\rho_{max}-\frac{1}{2}K\pi+\frac{1}{4}\pi\right)\right]\right)\right\} e^{-\beta\left(\hbar^{2}/2m\right)k^{2}}\end{array}\label{eq:b6}\end{equation}
 which can be rewritten as \begin{equation}
\begin{array}{c}
\frac{3^{1/2}}{\pi\lambda_{T}}\int_{0}^{\infty}dk{\displaystyle \sum_{\nu}}\,(\mp)_{\nu}\,\left\{ \frac{1}{2k}\left(\cos\left[2\left(k\rho_{max}\right)\right]\left[\cos\left(2\overline{\delta}^{\nu}\left(k\right)\right)+1\right]\right.\right.\\
\left.\left.-\sin\left[2\left(k\rho_{max}\right)\right]\sin\left(2\overline{\delta}^{\nu}\left(k\right)\right)\right)\right\} e^{-\beta\left(\hbar^{2}/2m\right)k^{2}}\end{array}\label{eq:c5}\end{equation}
where, as before,  $\mp$ takes the value of $-1$
when the K are even and $+1$ when they are odd. \\
 We see that, just as in the previous oscillating terms for the 1st
term of $b_{3}$, again, the cosine term poses problems and the sine
term does not contribute to the fugacity coefficient.

\newpage{}

\section*{Distorted Born}

From the foregoing, we see that to obtain analytical results for $b_{3}-b_{3}^{0}$,
we shall need an expansion of the eigenphase shifts, for small values
of $k$, our customary wave number.

The `conventional' result for a 2-body potential, as stated by Mott
and Massey \cite{mott}, is that given an orbital angular momentum $\ell\hbar$,
and a radial potential that falls off as $r^{-s}$ for large $r$,
then

\[
\lim_{k\rightarrow0}\, k^{2\ell+1}\,\cot(\eta_{\ell})=constant\]
 if $\ell<(s-3)/2$ \, . This is the usual result, in three dimensions,
for short range potentials.

\noindent If, however, $\ell>(s-3)/2$, Mott and Masey then obtain
\[
\lim_{k\rightarrow0}\, k^{s-2}\,\cot(\eta_{\ell})=constant\]
 a very different behaviour, that they show can be obtained by a Born
approximation. The point is that, here, it is the long-tail behaviour
of the potential that drives the low-energy behaviour of the phase
shifts.

In our hyperspherical formalism we see, for repulsive delta-function
interactions, from our equations (\ref{eq:int1}), that when $\rho$
is large, the role of the $\ell+1/2$ is played by $K+3$ or $K+1$,
respectively and, therefore, the criterion for the dominance of the
`tail' becomes $(K+3)>(s-2)$ for the $(123)$ case and $(K+1)>(s-2)$
when discussing the $(12,3)$ physical problem. We note that this
is satisfied for our adiabatic potentials.

Still, the recommended use of a Born approximation is not so straightforward.
This is due to a peculiarity arising from working in one dimension.
In a manner not arising in other dimensions, for any of our equations,
the \textit{effective} centrifugal term \textit{changes} as a function
of $\rho$, due to the interaction. Thus, for the $(123)$ case and
a given $K$, for `low' values of $\rho$ it assumes the centrifugal
value of $(K^{2}-1/4)/\rho^{2}$, while for `large' values of $\rho$,
thanks to the interactions, it takes on the asymptotic value of $(K+3)^{2}/\rho^{2}$.

Thus a `conventional' Born iteration, with the `low-$\rho$' expression
for the centrifugal term, and using $J_{K}(k\rho)$ as a zeroth order
approximation for the solution, would yield iterates that would not
form an expansion in powers of $k$, but would in fact each contribute
to a constant term in a power series expansion of an R-matrix. {[}Here,
to make the point, we just discuss one equation.]

We were then led to the development of the `{\em Distorted Born}'
\cite{larsen1,popiel0,popiel}.

For ample discussion, and details, see J.J. Popiel's thesis\cite{popiel0}.
Below, we follow his general outline, for the case $(123)$, which
we have extended to include $Ks$ that are odd, and also to the physics
that we denote by $(12,3)$. We are happy to have recovered and extended
his results; we simply note that we differ by a few signs (which,
in some crucial equations, compensate!).

For low energies, we write: \begin{equation}
\left[\frac{d^{2}}{d\rho^{2}}\;-\;\frac{(K+3)^{2}-\frac{1}{4}}{\rho^{2}}\;+\; k^{2}\right]\;\phi_{K}^{K^{\prime}}\left(k,\rho\right)\;\;=\;\;\sum_{K''}\;(\;\;\;\;)^{\infty}\;\;\phi_{K''}^{K^{\prime}}\left(k,\rho\right)\label{eq:mon}\end{equation}
 where \begin{equation}
(\;\;\;\;)_{K}^{K''}\left(\rho\right)=\left(\Lambda_{K}\left(\rho\right)\;-\;\frac{(K+3)^{2}}{\rho^{2}}\right)\delta_{K}^{K''}\;-\; D_{K}^{K''}\left(\rho\right)\;-\;2C_{K}^{K''}\left(\rho\right)\frac{d}{d\rho}\;\;\;.\label{eqmon}\end{equation}
The superscript $\infty$ on $(\;\;\;)^{\infty}$ is meant to imply
that we seek to use the large $\rho$ expansion of $(\;\;\;\;)_{K}^{K''}$.
Asymptotically, it will behave as $1/\rho^{3}$. The matrix C, which
only has off-diagonal elements, is proportional to $1/\rho^{2}$,
but only appears in the coupled equations as $C\;\frac{d}{d\rho}$
and hence for large $\rho$ is effectively proportional to $k/\rho^{2}$.
The $K^{\prime}$ denotes the $K^{\prime}$th solution.

An iteration of our coupled equations will yield a power series in
k with each new term in the successive iteration beginning with a
higher power of k.

We now write an integral equation, valid for small values of $k$:
\begin{equation}
\phi_{K}^{K'}(k,\rho)=(\phi_{K}^{K'})^{0}(k,\rho)\;\;+\;\;\int_{0}^{\infty}\; G_{K+3}(\rho,\rho';k)\;\;\sum_{K''}\;(\;\;\;\;)^{\infty}\;\phi_{K''}^{K'}(k,\rho^{\prime})\; d\rho'\end{equation}
 where $(\phi_{K}^{K'})^{0}$ is the leading term in an expansion
of $\phi_{K}^{K^{\prime}}$ for low energy. Our choice is \begin{equation}
(\phi_{K}^{K'})^{0}\left(k,\rho\right)=\delta_{K}^{K'}\;\sqrt{k\rho}\; J_{K+3}(k\rho)\end{equation}
 and%
\footnote{Change of sign from Popiel's result%
} \begin{equation}
G_{K+3}(\rho,\rho';k)=\left\{ \begin{array}{ll}
+\frac{\pi}{2k}\;\sqrt{k\rho}\; J_{K+3}(k\rho)\sqrt{k\rho'}\; N_{K+3}(k\rho') & \rho<\rho'\\
\; & \;\\
+\frac{\pi}{2k}\;\sqrt{k\rho}\; N_{K+3}(k\rho)\sqrt{k\rho'}\; J_{K+3}(k\rho') & \rho>\rho'\end{array}\right.\;\;\;.\end{equation}
 With this, the large $\rho$ behaviour of $\phi_{K}^{K^\prime}$ is \begin{equation}
\phi_{K}^{K'}\left(k,\rho\right)=\delta_{K}^{K'}\;\sqrt{k\rho}\; J_{K+3}(k\rho)\;+\; W_{K}^{K'}(k)\;\sqrt{k\rho}\; N_{K+3}(k\rho)\end{equation}
where$^{1}$ 
\begin{equation}
W_{K}^{K'}(k)=+\frac{\pi}{2k}\;\int_{0}^{\infty}\;\sqrt{k\rho}\; J_{K+3}(k\rho')\;\;\sum_{K''}\;(\;\;\;)^{\infty}\;\phi_{K''}^{K'}\left(k,\rho\prime\right)\; d\rho\prime \;\;\;.\end{equation}
 We note that the matrix W now contains all of the scattering information
beyond the ($k\rightarrow0$) result, and completely characterizes
the asymptotic form of the wave-function..

Iterating once, gives us: \begin{equation}
W_{K}^{K'}(k)=1st\; Distorted\; Born\;\;+\;\;\ldots\label{eq:102}\end{equation}
 where$^{1}$ \begin{equation}
\begin{array}{lll}
1st\; Distorted\; Born & = & +\left(\frac{\hbar^{2}k}{mg}\right)\;\left\{ \frac{24\sqrt{2}}{\pi}\frac{(K+3)(K'+3)}{[(K+K'+6)^{2}-1]\;[(K-K')^{2}-1]}\right\} \\
\; & \; & \;\\
\; & \; & +\delta_{K}^{K'}\times\left(\frac{\hbar^{2}k}{mg}\right)^{2}\left\{ \frac{\left(\frac{27}{\pi}+\frac{\pi}{12}\right)\;(K+3)}{(K+2)(K+4)}\;+\;\frac{9}{4\pi(K+2)(K+3)(K+4)}\right\} \\
\; & \; & \;\\
\; & \; & +\; O(k^{3})\end{array}\end{equation}

The 2nd \textit{Distorted Born} yields a $k^{2}$ contribution which
exactly cancels that obtained from the 1st Born! So, the total contribution
of order $k^{2}$ is zero. \\

\noindent For the case $(12,3)$, and $\overline{W}^{K^\prime}_{K}$, 
the corresponding formula is: 
\begin{equation}
\begin{array}{lll}
1st\; Distorted\; Born & = & +\left(\frac{\hbar^{2}k}{mg}\right)\;\left\{ \frac{8\sqrt{2}}{\pi}\frac{(K+1)(K'+1)}{[(K+K'+2)^{2}-1]\;[(K-K')^{2}-1]}\right\} \\
\; & \; & \;\\
\; & \; & +\; O(k^{2})\end{array}\end{equation}
 The terms of $O(k^{2})$ are very much more difficult to evaluate,
as the individual diagonal terms for $K=0$ diverge! As we can see
from Mott and Massey's discussion we are, for $K=0$ (and $s>3$,
corresponding to higher inverse powers appearing also in the expansion
of the interactions at large distances), at the limit of the domain
of validity of this method of calculation.

In Appendix D we show for a simplified version of a model mentioned
in K. Chadan and P.C. Sabatier's book \cite{chadan}, 
for which the exact solution is known, that our \textit{Distorted
Born} approach gives exactly the correct coefficients, where the Mott
and Massey criterion states that it should (and which is also precisely
where the integrals of the expansion converge).

\newpage{}

\section*{The Phase Shift Sums}

\subsection*{from $W_{K}^{K}$}

From Eqs. $(\ref{eq:b})$ and $(\ref{eq:b5})$,
we see that we need to evaluate 
\begin{equation}
\sum_{\nu}\left[\frac{d}{dk}\delta^{\nu}(k)\right]-\sum_{\nu'}
\left[\frac{d}{dk}\overline{\delta}^{\nu'}(k)\right].
\label{diff1}
\end{equation}
 To do so, we note that our phase shifts are those obtained from 
diagonalizations of our matrices of $W$  and $\overline{W}$, respectively.
In each case, the sum of the tangents of their phase shifts forms
a trace which is, of course, invariant under changes in the representation.
\\
Thus we can, for example, write 
\begin{equation}
\sum_{\nu}\left[-\tan(\delta_{W}^{\nu})\right]=Trace(W)=
\sum_{K}W_{K}^{K}\label{diff2}
\end{equation}
 but since we expand the phase shift in powers of $k$, and keep at
most the first two terms , we find that we can evaluate Eq.(\ref{diff1})
in first order {\em{Distorted Born}}, as: \begin{equation}
\frac{\hbar^{2}}{mg}(\frac{24\sqrt{2}}{\pi})\,\left[\sum_{0,3,6...}\;\frac{(K+3)^{2}}{4(K+3)^{2}-1}-\frac{1}{3}\sum_{0,1,2...}\;\frac{(K+1)^{2}}{4(K+1)^{2}-1}\right],\label{DBdiff}\end{equation}
 where we have combined the {\em cosine} and the {\em sine}
contributions. \\
 As we can see, and as is appropriate for our clusters, the individual
contributions from the $(123)$ and the $(12,3)$ terms are infinite!
We expect the difference to be finite. \\

\noindent
Let us comment: \\
As shown in a semi-classical treatment\cite{larsen, larsen2}, each of the
phase-shift sums corresponding to the above, in 3 dimensions, is associated 
with a classical expression integrand, which integrated over the      
position variables, diverges as the volume tends to infinity.
When all of the terms of the cluster are taken together, the resulting
integral is finite and volume independent.

\subsection*{Abel summation}

We see, therefore, that the infinities are the result of taking the 
infinite volume limit ... and the differences of the phase shift sums
must be taken with great care.  \\

The literature of the evaluation and manipulation of divergent series 
is rich and of great intellectual interest. 
We give some references\cite{hardy}. \\

A famous method to evaluate such series is due to Abel. 
Using his method we introduce an additional parameter `$x$', and powers of
$x$,  which will allow us to obtain finite results for each of our 
divergent sums (for $0 < x <1$), subtract our sums, 
let $x = 1$ and then revert to our `physical' results. \\

We first start by rewriting the sums in (\ref{DBdiff}) as 
\[
\sum_{3,6...}^{\infty}\ \frac{K^{2}}{4\ K^{2}-1}-\frac{1}{3} \, 
\sum_{1,2...}^{\infty}\ \frac{K^{2}}{4\ K^{2}-1}\]
Following Abel, we multiply the  
terms of the diverging sums by the strongly decreasing function $x^{K}$
where $0<x<1$ 
\[
\sum_{3,6...}^{\infty}x^{K}\ \frac{K^{2}}{4\ K^{2}-1}-\frac{1}{3}\ \sum_{1,2...}^{\infty}x^{K}\ \frac{K^{2}}{4\ K^{2}-1}.\]
 Here we see that for equal values of $K$, on both sides, say $K=3$,
the same term appears in both sums, multiplied by $x$ to the same
power. This is important. \\
Changing the index in the first sum, we obtain 
 \[
\sum_{1,2...}^{\infty}x^{3K}\ \frac{K^{2}}{4K^{2}-1/9}-\frac{1}{3}\ \sum_{1,2...}^{\infty}x^{K}\ \frac{K^{2}}{4\ K^{2}-1}\]
For the first term, then, 
since the series is absolutely convergent
for $0 <x <1 $, we can write:
\begin{equation}
\frac{1}{4}\sum_{0,3,6...}^{\infty}x^{K+3}+\ \frac{1}{16}
\sum_{0,3,6...}^{\infty}x^{K+3}\ \left(\frac{1}{(K+3)^{2}-\frac{1}{4}}
\right)
\end{equation}   
Similarly, for the second sum, involving the $(12,3)$ terms, we
write: 
\begin{equation}
\ \frac{1}{4}\sum_{0,1,2...}^{\infty}x^{K+1}+\ \frac{1}{16}
\sum_{0,1,2...}^{\infty}x^{K+1}\ \left(\frac{1}{(K+1)^{2}-\frac{1}{4}}\right)
\end{equation}
 The expression in the bracket in (\ref{DBdiff}) then becomes: 
\begin{equation}
\frac{1}{4}\left(\sum_{0,3,6...}^{\infty}x^{K+3}-\frac{1}{3}
\sum_{0,1,2...}x^{K+1}\right)+\frac{1}{16}
\left(\sum_{0,3,6...}^{\infty}x^{K+3}\ \frac{1}{(K+3)^{2}-\frac{1}{4}}-
\frac{1}{3}\ \sum_{0,1,2...}^{\infty}x^{K+1}\ \frac{1}{(K+1)^{2}-
\frac{1}{4}}\right)
\end{equation}
 Collecting the divergent parts, gives us 
\begin{equation}
\sum_{0,3,6...}^{\infty}x^{K+3}-\frac{1}{3}\sum_{0,1,2...}x^{K+1}=
\frac{x^{3}}{1-x^{3}}-\frac{1}{3}\ \frac{x}{1-x}=
-\frac{x+2x^{2}}{3(1+x+x^{2})}
\end{equation}
 which tends to $-1/3$ when $x\to1^{-}$\, . \\
 On the other hand 
\[ \sum_{0,3,6...}^{\infty}\frac{1}{(K+3)^{2}-\frac{1}{4}}=
\frac{1}{9} \sum_{1,2,3...}^{\infty}\frac{1}{j^{2}-\frac{1}{36}}=
2-\frac{\pi}{\sqrt{(}3)}\]
 and \[ \sum_{0,1,2...}^{\infty}\frac{1}{(K+1)^{2}-\frac{1}{4}}=
\sum_{1,2,3...}^{\infty}\frac{1}{j^{2}-\frac{1}{4}}=2\]
 So that we obtain for (\ref{DBdiff}) 
\begin{equation}
\frac{\hbar^{2}}{mg}\frac{24\sqrt{(}2)}{\pi}
\left\{ -\frac{1}{4}\times\frac{1}{3}+\frac{1}{16}
\left( 2- \frac{\pi}{\sqrt(3)} - \frac{1}{3}\; 2 \right)
\right\} =-\frac{\hbar^{2}}{mg}\ \frac{\sqrt{6}}{2}
\label{pdresult}
\end{equation}

\newpage
\subsection*{The oscillatory terms}

\textbf{Abel summation applied to the sums for the oscillating terms.}
 \\

\noindent 
Let us use the Abel procedure for the elements of the sums
that appear in the integrands of the oscillating terms. I.e., recall
that in (\ref{eq:c}) and (\ref{eq:c5}) we obtained a term such as 
$(\mp)^{\nu}[\cos(2\delta^{\nu}(k))+1]$,
which gives a $-2$ for the even $K$ terms and a $2$ for the odd
$K$ terms, as $k\rightarrow0$. Using, then a `trace' argument, we
pass from the sums over $\nu$ to sums over the indices $K$ associated
with our solutions. \\
 I.e., if we take one of the $W$ - associated with one of the sums
in the oscillatory terms - say for a $(123)$, $even$ basis - and
diagonalize it, we will then obtain $\delta^{K}$ for a set of values
of $K$, such as $0,6,12,...$, etc. We can use these values of $K$
to be our values of $\nu$, \, i.e. to be the values of the index
over which we sum. \\
 We will then have \[
\sum_{0,6,12...}^{\infty}x^{K+3}-\sum_{3,9,15...}^{\infty}x^{K+3}-\sum_{0,2,4...}x^{K+1}+\sum_{1,3,5...}x^{K+1}\]
 which gives 
\[
\frac{x^{3}\ (1-x^{3})}{1-x^{6}}-\frac{x(1-x)}{1-x^{2}}=\frac{x^{3}}{1+x^{3}}-\frac{x}{1+x}\]
 then \[
\frac{x^{3}}{1+x^{3}}-\frac{x(1-x+x^{2})}{1+x^{3}}=\frac{x(x-1)}{1+x^{3}}\]
 which goes to $0$ as $x\rightarrow1$. \\

 Further, if in the integral one makes the change of variable $u=k\rho_{max}$,
then when one takes the limit $\rho_{max}\rightarrow\infty$, the
argument of the $\delta^{\nu}(u/\rho_{max})$ will go to zero, and
if we take the appropriate sums we find ourselves with the same expressions
that we have evaluated above. The integrand of the total oscillatory
contribution will be exactly zero.

\newpage{}

\section*{The Fugacity Coefficient}

\subsection*{our result}

Our fundamental formula is: \begin{equation}
b_{3}-b_{3}^{0}=\frac{3^{1/2}}{\lambda_{T}}\int_{0}^{\infty}dk\, e^{-\beta\,\left(\hbar^{2}/2m\right)\, k^{2}}\frac{1}{\pi}\left(\sum_{\nu}\frac{d}{dk}\delta^{\nu}(k)-\sum_{\nu\prime}\frac{d}{dk}\overline{\delta}^{\nu\prime}(k)\right)\end{equation}
 Putting the values that we have just obtained (\ref{pdresult}) for
the derivative of the difference of the phase shift sums, we obtain:
\begin{equation}
\frac{3^{1/2}}{\lambda_{T}}\left(-\frac{1}{\pi}\frac{\hbar^{2}}{mg}\frac{\sqrt{6}}{2}\right)\,\int_{0}^{\infty}dk\, e^{-\beta(\frac{\hbar^{2}}{2m})k^{2}}\end{equation}
 and remembering that $\lambda_{T}$ can also be written as $(2\pi\hbar^{2}\beta/m)^{1/2}$
we obtain our result: \begin{equation}
b_{3}-b_{3}^{0}=-\frac{3\sqrt{2}}{4\pi g}\,\frac{1}{\beta},\label{fresult}\end{equation}
 which agrees with the first term of the expansion of the integrals
of Dodd and Gibbs\cite{dodd}. \\
 See Appendix C.

\newpage
\section*{Conclusion}
Our aim, our hope, is to evaluate, express the virial coefficients
in terms of scattering, asymptotic quantities of the wave functions, 
and bound state energies. 

\esp
This was done in the 1930's for the 2nd virial by 
Uhlenbeck and Beth\cite{uhlenbeck}, and 
by Gropper\cite{Gropper}.
Our dream has been to do the same thing for the `higher' virials.
We believe that we have made progress in doing so for all
the `higher' virials - since our approach can be generalized to accommodate 
any number of particles - but demonstrate our method for a prototype third 
fugacity cluster, required for the 3rd virial. 

\esp
In this paper, 
we present results for a model in 
\mbox{1-dim.} (with delta function interactions) 
because this model has attracted the attention of many 
workers\cite{McGuire, yang, Dodd}, 
due to its simplicity, 
which allows the evaluation of many results, expressions, and steps, 
analytically.
Further, Dodd and Gibbs\cite{dodd}, uniquely, were able to look at 
the Statistical Mechanics and evaluate the 
third fugacity cluster in \mbox{1-dim.} and we anticipated that we would 
be able 
to compare with their result.
 
\esp
In essence, our overall reasoning is simple and, we think, transparent! 

\esp
Our fugacity clusters are expressed in terms of traces of terms such
as $Tr(e^{- \beta H_{n}})$.
We know that we can introduce complete sets of position eigenstates
in these traces, which will then require us to integrate over position
variables, and a wave number coordinate, as done for our model 
by Dodd and Gibbs\cite{dodd}.

\esp
We do introduce complete sets of states, which we choose 
to expand in a hyperspherical adiabatic basis,
and then integrate {\bf analytically} over all position 
coordinates - including our hyper-radius coordinate $\rho$ - 
obtaining results characterized, here,  
by the asymptotic behaviour of the wave functions (i.e. for large $\rho$), 
described by {\bf eigenphase shifts}, functions of the wave number.

\esp
This approach is general. Yes, there are subtle problems involving bound 
states, but here, in our 1-dim. repulsive case, we need not face them.
In 1-dim., however, we find that there is a possibility of a 
contribution from oscillating terms, which for example do 
contribute\cite{2nd} to 
$b_2$, in the case of Bose statistics. Our results indicate that there 
is no such contribution for $b_3$ in Bose statistics.

\esp
A delicate problem that we face, here, is the subtraction between 
phase shift sums. As we have shown, after evaluating the c.m. contribution, 
we must subtract contributions proportional to $L^2$, so as to obtain 
$L$-independent results. 
In our formalism, the first subtraction takes place, eliminating 
one power of $L$ and yielding 2 types of W-phase shifts. We then saw that 
we had to subtract these two phase shifts sums to yield a finite, 
i.e. L-independent, answer. 
Using the results only of order $k$, we were successful in doing so, 
using a summation procedure associated with Abel.

\esp
For our 3 particles in one dimension, subject to delta-function interactions,
we were able, for the fully interacting case (123), using a \textit{distorted 
Born} approximation, to obtain the phase shifts (or their tangents), to order
$k^2$. In our hyperspherical adiabatic formalism, they are associated with 
the `tail' of long ranged potentials. 

\esp
For the (12,3) case, 
we encountered reliable results for the lowest contribution,
to order $k$. However, for $K = 0$, and the coefficient of $k^2$,  we obtain  
divergences from the contributing terms. 
Added together, these divergences cancel but yield a
finite term that we cannot trust. See, in Appendix D, our analysis of the 
domain of applicability of the \textit{distorted Born}. 
[We note that we tried - very hard, in many ways, using different methods - 
to obtain the coefficient 
 of the $k^2$ term, but could not.]

\esp
Using, then, the result of order $k$, which we do believe is correct in our
\textit{distorted Born} formalism,  we obtain a result, 
for the temperature  behaviour 
of $b_3$, 
which agrees with the leading term in an expansion of the double integral
proposed by Dodd and Gibbs.

\esp

We feel justified, therefore, in believing in the validity of our formalism,
and of its extension to 1-dim. and Bose particles.

\esp

Ultimately, in our future plans, we will wish to extend, rigorously, 
our general formalism, to include cases of three-body systems  with 
attractive potentials and bound states. Specifically,
we need to be able to accommodate states in which a possible outcome,
asymptotically,  would involve, for example,  
a 2-body bound state + an asymptotically free particle.
In an old paper\cite{larsen}, we proposed a formula to do this, 
a very intuitive formula, without a proper proof.  
Surely, one of our aims must to be 
remedy that! \\
(We note that to take into account a 3-body bound state is trivial;
in the evaluation of the trace we simply add an exponential term
involving the bound state energy.)

\newpage{} 
\section*{Acknowledgments}
A. Amaya and S. Y. Larsen acknowledge Direcci\'on General de Asuntos 
del Personal Acad\'emico for partial support under project 
No. \mbox{IN-105707-3.} S. Y. Larsen also gratefully thanks the
Instituto de Ciencias F\'{i}sicas and the Institut de Physique Nucl\'eaire 
for their always splendid hospitality.

\newpage

\section*{Appendix A}

See the discussion in the article by Larsen and Poll \cite{poll}.
\\

\noindent \begin{equation}
{\phi}_{j}^{i}\left(k,\rho\right)\sim\left(k\rho\right)^{1/2}(\delta_{j}^{i}\, J_{j}\left(k\rho\right)+R_{j}^{i}(k)\, N_{j}\left(k\rho\right))\label{fia1}\end{equation}

\noindent As the reactance matrix $\left\Vert R_{j}^{i}\right\Vert $
is real and symmetric, it can be diagonalized by a real orthogonal
matrix $\left\Vert \mathcal{C}_{j}^{i}\right\Vert $. Consequently,
we can introduce eigensolutions: \begin{equation}
\phi_{j}^{\nu}\left(k,\rho\right)\sim\left(k\rho\right)^{1/2}\mathcal{C}_{j}^{\nu}\left(J_{j}\left(k\rho\right)+\Lambda_{\nu}(k)\, N_{j}\left(k\rho\right)\right)\label{fia2}\end{equation}
 characterized by a unique eigenvalue for all of its amplitudes.

\noindent The eigenphase shifts are then defined by: 
\begin{equation}
\Lambda_{\nu}(k) =-\tan\delta_{\nu}(k)
\label{ps}\end{equation}

\newpage
\section*{Appendix B}

In this Appendix we give the asymptotic expression of the interaction
$\left(\; \; \; \right)_{K}^{K'}$ entering the coupled equations (Eqs.
{(}\ref{eq:mon},\ref{eqmon})). It consists of a series in terms of
$1/\rho$ and is used in our first and second Born approximations.
We put for the $\left(123\right)$ case \begin{equation}
q_{K}=(K+3)+\sum_{j}\frac{a_{j}}{\rho^{j}}\label{dev3}\end{equation}
 with $K = 0$ modulo $6$ \, for $K$ even,  and $K = 3$ modulo $6$ \,
for $K$ odd, and for the $2+1$ case \begin{equation}
q_{K}=(K+1)+\sum_{j}\frac{{\bar{a}_{j}}}{\rho^{j}}\label{dev1}\end{equation}
 with $K = 0$ modulo $2$ \, for $K$ even, and $K = 1$ modulo $2$ \,
for $K$ odd. This leads to (see the previous discussions on the 
calculations of the adiabatic eigenfunctions)
\begin{equation}
q_{K}=(K+3)-36\left (\frac{K+3}{c\pi^{2}\rho}\right )+1296\left (\frac{K+3}
{c^{2}\pi^{4}\rho^{2}} \right)+
{\cal O}\left(\frac{1}{\rho^{3}}\right)\end{equation}
 for every value of $K$ in the $(123)$ case and

\begin{equation}
q_{K}=(K+1)-12\left(\frac{K+1}{c\pi^{2}\rho}\right )+
144\left (\frac{K+1}{c^{2}\pi^{4}\rho^{2}}\right)+
{\cal O}\left(\frac{1}{\rho^{3}}\right)\end{equation}
 for every value of $K$ in the $(12,3)$ case. \\
In both cases  $c=(2m/\hbar^{2})(3g/\pi\sqrt{2})$. \\

Obtaining the asymptotic expansion of $\Lambda_{K},\ C$ and $D$ matrices
(See definitions in Eqs. {(}\ref{ang},\ref{Cs},\ref{Ds})) in terms
of $1/\rho$, we recover (and extend) the results for $\left(123\right)$
of Larsen and Popiel \cite{popiel}.:

\begin{equation}
\Lambda_{K}(\rho)=\left(q_{k}^{2}-1/4\right)/\rho^{2}=\frac{(K+3)^{2}-1/4}{\rho^{2}}+{\cal O}\left(\frac{1}{\rho^{3}}\right)\label{diag3}\end{equation}

\begin{eqnarray}
C_{K}^{K'}(\rho) & = & -\frac{d}{\rho^{2}}\ \frac{12\sqrt{2}\ (K+3)(K'+3)\cos(\pi(K-K')/6)}{(K-K')(K+K'+6)\ \pi}\nonumber \\
 & + & \frac{d^{2}}{\rho^{3}}\
\frac{144(K+3)(K'+3)\cos(\pi(K-K')/6}{(K-K')(K+K'+6)\
\pi^{2}}+{\cal O}\left(\frac{1}{\rho^{4}}\right)\end{eqnarray}

and

\begin{eqnarray*}
D_{K}^{K'}(\rho) & = & \frac{d}{\rho^{3}}\ \frac{24\sqrt{2}(K+3)(K'+3)\cos(\pi(K-K')/6)}{(K-K')(K+K'+6)\pi}\\
 & - & \frac{d^{2}}{\rho^{4}}\ \frac{144(K+3)(K'+3)(5(K+3)^{2}-(K'+3)^{2})\cos(\pi(K-K')/6)}{(K-K')^{2}(K+K'+6)^{2}\pi^{2}}+{\cal O}\left(\frac{1}{\rho^{5}}\right)\end{eqnarray*}

and for $K=K'$

\begin{equation}
D_{K}^{K}(\rho)=-\frac{d^{2}}{\rho^{4}}\ 
\frac{2(27+(K+3)^{2}\pi^{2})}{3\pi^{2}}+{\cal O}\left(\frac{1}{\rho^{5}}\right)
\end{equation}

\noindent
[Note that $C_{K}^{K'}$ is an antisymmetric matrix and that, therefore, the
diagonal elements are zero.]  \\

\esp
\noindent
In these expressions $d = \hbar^2/mg$.
\newpage
\noindent We obtain in the new case, $\left(12,3\right)$,

\begin{equation}
\Lambda_{K}(\rho)=\frac{(K+1)^{2}-1/4}{\rho^{2}}+{\cal O}\left(\frac{1}{\rho^{3}}\right)\label{diag12}\end{equation}

\begin{eqnarray}
C_{K}^{K'}(\rho) & = & -\frac{d}{\rho^{2}}\ \frac{4\sqrt{2}\ (K+1)(K'+1)\cos(\pi(K-K')/2)}{(K-K')(K+K'+2)\ \pi}\nonumber \\
 & + & \frac{d^{2}}{\rho^{3}}\
\frac{16(K+1)(K'+1)\cos(\pi(K-K')/2)}{(K-K')(K+K'+2)\
\pi^{2}}+{\cal O}\left(\frac{1}{\rho^{4}}\right)\end{eqnarray}

and

\begin{eqnarray*}
D_{K}^{K'}(\rho) & = & \frac{d}{\rho^{3}}\ \frac{8\sqrt{2}(K+1)(K'+1)\cos(\pi(K-K')/2)}{(K-K')(K+K'+2)\pi}\\
 & - & \frac{d^{2}}{\rho^{4}}\ \frac{16(K+1)(K'+1)(5(K+1)^{2}-(K'+1)^{2})\cos(\pi(K-K')/2)}{(K-K')^{2}(K+K'+2)^{2}\pi^{2}}+{\cal O}\left(\frac{1}{\rho^{5}}\right)\end{eqnarray*}

for $K\ne K'$, and for $K=K'$ we have

\begin{equation}
D_{K}^{K}(\rho)=-\frac{d^{2}}{\rho^{4}}\ \frac{2(3+(K+1)^{2}\pi^{2})}{3\pi^{2}}+{\cal O}\left(\frac{1}{\rho^{5}}\right)\end{equation}

\newpage{}

\section*{Appendix C}

\noindent \textbf{Expansion of the integrals of Dodd and Gibbs} \\

\noindent For their concluding equation Eq.(18), in the main part
of their paper, Dodd and Gibbs\cite{dodd} write: \begin{equation}
b_{3}-b_{3}^{0}=\frac{6}{(2\pi)^{3}}(\frac{\pi}{\beta})^{3/2}\int_{0}^{\infty}ds\int_{0}^{\infty}dt\, e^{-(1/2\beta)(s^{2}+t^{2}+st)}\,(e^{-cs-ct}-1)\, e^{-cs}\end{equation}
 Denoting, the integral above as ${\cal I}$, the integral can be
rewritten as \[
2\beta\int_{0}^{\infty}d{s}\int_{0}^{\infty}d{t}\exp{(-({s}^{2}+{t}^{2}+{s}{t}))}(\exp{(-\lambda(s+t))}-1)\exp{(-\lambda s)},\]
 where we have set $\lambda=c\sqrt{2\beta}$, and written again as:
\\
 \[
\;\;\;\;(2\beta)\,\frac{\sqrt{\pi}}{2}\int_{0}^{\infty}ds\, e^{-s^{2}}\{e^{-2\lambda s+\frac{(\lambda+s)^{2}}{4}}\, erfc\left(\frac{\lambda+s}{2}\right)-e^{-\lambda s+\frac{s^{2}}{4}}\, erfc\left(\frac{s}{2}\right)\}\]
 This then leads to the following expansion: \[
{\cal I}/(2\beta)\,\simeq\,-\frac{\sqrt{\pi}}{2\ \lambda}+\frac{1}{\lambda^{2}}+\frac{3\ \sqrt{\pi}}{4\ \lambda^{3}}-\frac{4}{\ \lambda^{4}}+{\cal {O}}\left(\frac{1}{\lambda^{5}}\right)\]
 Taking the first term, we obtain: \\
 \begin{equation}
b_{3}-b_{3}^{0}=\frac{6}{(2\pi)^{3}}(\frac{\pi}{\beta})^{3/2}\,(2\beta)\,\left(-\frac{\sqrt{\pi}}{2c\sqrt{2\beta}}\right)=-\frac{3}{4\pi\sqrt{2}}\,\frac{1}{\beta c}\end{equation}
 Looking then at the two papers, we determine that $\beta$ of DG
equals the $(\hbar^{2}/2m)\,\beta$ that we use, and that the $2c$,
used above, is our $g\,(2m/\hbar^{2})$. Substituting then, in their
result, gives our formula: \[
b_{3}-b_{3}^{0}=-\frac{3}{4\pi\sqrt{2}}\,\frac{1}{g\beta}.\]

\newpage

\section*{Appendix D}

\noindent \textbf{A little model, testing {\em Distorted Born}}
\\

\noindent Consider the following differential equation: 
\begin{equation}
\left(-\frac{d^{2}}{dr^{2}}+\frac{2a^{2}}{(1+ar)^{2}}\right)\psi(k,r)
=k^{2} \, \psi(k,r).
\end{equation}
 I.e., a Schr\"{o}dinger equation with a curious potential which, while
finite at the origin, tends to $2/r^{2}$ at large distances. Thus,
it mimics the situation that we have encountered in this work, in
which the `effective' centrifugal term has changed as a function of
distance. \\
 The exact solution is: 
\begin{equation}
\psi(k,r)=
\frac{-a^{2}(kr)\cos(kr)+(a^2+k^{2}+ak^{2}r)\sin(kr)}{k^{2}(1+ar)}
\end{equation}
 which, for $r$ large, becomes proportional to \[
\sqrt{kr}\,\left[\frac{a}{k}\, J_{3/2}(kr)-N_{3/2}(kr)\right]\]
 which implies that, introducing a $W$ and a phase shift $\delta$,
we obtain an exact result: $W=-k/a$, and therefore $\tan(\delta)=k/a$.
There are no terms in $k^{2}$, or $k^{3}$ for the tangent of $\delta$.
\\

For $r$ large, the potential takes the form $2/r^{2}-(4/a)1/r^{3}+(6/a^{2})1/r^{4}...$
\, and using the $-(4/a)1/r^{3}$ in our first {\em distorted Born}
yields \begin{equation}
\frac{\pi}{2k}\int_{0}^{\infty}dr\,\sqrt{kr}J_{3/2}(kr)\,\sqrt{kr}J_{3/2}(kr)\,\frac{(-4/a)}{r^{3}}\,,\end{equation}
 which equals \begin{equation}
-\frac{2\pi}{a}\, k\int_{0}^{\infty}dz\,\left(J_{3/2}(z)\right)^{2}\frac{1}{z^{2}}=-\frac{k}{a}.\end{equation}
 We note that the integrand behaves as $z$ at the origin - hence
presents no difficulty there - and so would a term in the potential
of $1/r^{4}$, which would give $1/z^{3}$ and an integrand which
would be finite at the origin. A term in the potential of $1/r^{5}$
would lead to a divergence. \\

If we use our {\em distorted Born} formulation to 2nd order, we
obtain : \[
W=-k/a+(k/a)^{2}\,(2\pi/5-2\pi/5)=-k/a\]
where the 1st $2 \pi/5$ comes from the $1/r^4$ term in the 1st Born,
and the 2nd $2 \pi/5$ from the 2nd Born. \\

We see that in this little model, when the integrals converge, 
we obtain the correct answer to order $k^2$. 
It is the $k^3$ term in this model (for W) which plays the role of the
$k^2$ term in our (2+1) case. I.e. in these cases [model and (2+1)]
the relevant integrals diverge. \\

We note that the criterion, given by Mott and Massey for the dominance of the
tail in determining the low $k$ behaviour, is that $\ell > (s-3)/2$, where 
$s$ characterizes the inverse power of r being considered. Here, our divergent
term is associated with $\ell = 1$ and $s = 5$. The inequality is thus 
violated and the Born (and also our distorted Born) has reached the limit 
of its application..

\end{document}